\documentclass[twocolumn, showpacs,preprintnumbers,amsmath,amssymb,prl]{revtex4}
\usepackage{amssymb}
\usepackage{amsfonts}
\usepackage{graphicx}
\usepackage{dcolumn}
\usepackage{bm}

\usepackage{times,epsfig,amssymb,amsmath}

\begin{document}

\title{Testing quantum entanglement with local measurement}

\author{Qing Xie$^{1,2}$\footnote{xieqingnbu@163.com}, X.-X. Wu$^2$,
X.-M. Ding$^3$, W.-L. Yang$^4$,
R.-H. Yue$^1$, and Heng Fan$^2$\footnote{hfan@iphy.ac.cn}}%
\affiliation{%
$^1$Faculty of Science, Ningbo University, Ningbo, 315211, China\\
$^2$Institute of Physics, Chinese Academy of Sciences, Beijing
100190, China\\
$^3$Institute of Applied Mathematics, Academy of Mathematics and
System Science, Chinese Academy of Sciences, Beijing 100190, China\\
$^4$Institute of Modern Physics, Northwest University, Xi'an 710069,
China }
\date{\today}

\begin{abstract}
We propose to detect quantum entanglement by a condition of local measurments.
We find that this condition can detect efficiently the pure entangled states for both
discrete and continuous variable systems.
It does not depend on interference of decoherence
from noise and detection loss in some systems, which allows a loophole-free test in
real experiments. In particular, it
is a necessary condition for the violation of some generalized Bell
inequalities.
\end{abstract}
\maketitle

\section{Introduction}
Quantum mechanics (QM) predicts correlations
that no local realistic theory can ever reproduce. 
Local realistic theory
states that there are hidden
variables which determine the properties of physical system prior to
and independent of measurement; also another assumption states that
these properties of one system can not influenced by one another
when these two systems are spacelike separated, which means
spacelike separated systems have their own hidden variables, i.e.
position and momentum in phase space.

Bell \cite{Bell,Bell2} proposed his theorem that QM violates certain
correlation function inequalities that any local hidden variable
(LHV) theory must satisfy. This finding directly displays
the incompatibility between QM and LHV theory for the first time.
After Bell's seminal work, multiple Bell-type inequalities have been
derived \cite{CHSH,Mermin1,CGLMP,ZB,gottesman2} and Bell inequality violations have been
observed experimentally \cite{CRSS,weihs,CRSS1,Brun}. Nowadays, it is well known that the
correlations QM imposed on outcomes of measurements carried out on
spacelike separated systems are strictly larger than LHV model does.


In quantum information processing, entanglement plays a key role and is the invaluable
resource. It is expected that
the entangled state shared by spatially separated parties will violate
some Bell inequalities \cite{gisin}.
On the other hand, if we restrict to pure states,
we may propose some equalities based on local measurements which is much simpler
than the Bell inequalities. These equalities is based on the fact that
a separable pure state is simply the product state, so local measurements
will not affect the measurements on other spatially separated parties.
In this paper, we will apply this method to study the entanglement. We remark
that this method is closely related with Bell inequality as we will show later,
it was used to derive the Bell inequalities by Mermin for multipartite entangled
states \cite{Mermin1}. Also it is closely related with contextuality \cite{ks,physreport},
as is well-known that QM is non-contextual, here our measurement
is performed locally.

If one of those equal relations is
infringed, we can conclude that the state under consideration is entangled.
In practice, to observe a
violation in real experiment, one may choose appropriate quantum
system and measurement settings to make sure the two sides of these
equality as different as possible to fight the interference of
docoherence. This method can be applied on both discrete and continuous variable
(CV) quantum systems. The application of CV case are not only of
importance from the fundamental point of view, but also may be a
applicable candidate to provide a loophole-free test. We present an
explicit example to demonstrate that these correlation function
equality
are effective both for discrete and continuous variable
system. Explicitly, any two qubit pure entangled state violate this
equality, for discrete spin operators measurement and continuous
quadrature operators measurement. The interference of decoherence
from noise and detection loss in real experiment always decrease the
possibility to observe a violation of Bell-type inequality which
increases the difficulty and complexity to design an experimental
scheme. In comparison, these interference from noise and detection
loss decrease both side of our correlation function equality at the
same rate. As we define the degree of violation by the ratio of two
side of these equality, this quantity is unaffect by docoherence
which make our equality particularly suitable for designing a
loophole-free test. Particularly, we demonstrate that the
general form of correlation function equality being violated is
responsible for the violation of Cavalcanti-Forster-Reid-Drummond
(CFRD) inequality, a Bell-type
inequality for CV system proposed recently \cite{CFRD}.
Thus experimentally there is no need to measure all the commeasurable
operators to observe a violation of CFRD inequality, only the
responsible term alone can lead to a conclusion.

The paper is organized as follows: In section II, we propose
the method and show that it can be applied to test entangled state.
In section III, we apply this method to CV system. In section IV, we
show that the violation of of those equality is responsible for the violation
of CFRD inequality. In section V, we extend the previous work
of CFRD inequality to more general case. Section VI is a brief conclusion.

\section{Correlation function equalities for separable pure states}
Let's first introduce the
Bell-type inequality and the notations. The simplest correlation
function inequality might be the Clauser-Horne-Shimony-Holt (CHSH)
Bell-type scenario \cite{CHSH}: two
space-separated observers, called Alice and Bob, each of them can
choose between two dichotomic observables of two outcomes, labeled
by $\pm1$. The CHSH inequality then reads
\begin{align}
E_{11}+E_{12}+E_{21}-E_{22}\leq 2,
\end{align}
where the functions $E_{ij}=\langle A_{i}B_{j}\rangle$, which are
expectation values of Alice's observable choice $A_{i}$ and Bob's
observable choice $B_{j}$, $i,j\in\{1,2\}$, known as correlation
functions. For convenience we also use $A_{i}$ $(B_{j})$ denoting
the outcome of Alice's (Bob's) measurement. It is well known that
(1) holds for any LHV model.
While in QM description, $A_{i}$ and
$B_{j}$ are measurement operators with eigenvalues $\pm1$ which can
generally be written as $A_{i}=\textbf{a}_{i}\cdot\bm\sigma$ and
$B_{j}=\textbf{b}_{j}\cdot\bm\sigma$, where
$\bm\sigma=(\sigma_{x},\sigma_{y},\sigma_{z})$ are Pauli matrices
and $\textbf{a}_{i},\textbf{b}_{j}$ are real unity vectors. Quantum
entangled state violates CHSH inequality. For example, choosing local
measurement $A_{1}=\sigma_{x}$, $A_{2}=\sigma_{y}$ for Alice and
$B_{1}=\sigma_{x}$, $B_{2}=\sigma_{y}$ for Bob, then CHSH inequality
is maximally violated when system is prepared in maximally entangled
state, $(|00\rangle +e^{i\phi }|11\rangle )/\sqrt {2}$.

One important assumption of LHV theory is that Alice's measurement
on her own system can not influence Bob's when these two systems are
spacelike separated.
Then the above four correlation functions
satisfy the following condition for any pure state with a product form, i.e.,
for a separable state,
\begin{eqnarray}
E_{ij}E_{kl}=E_{il}E_{kj}, \label{equality}
\end{eqnarray}
where $i,j,k,l\in\{1,2\}$. Here thus we propose to use those
equation as the detection of entangled states. Suppose system is
prepared as a pure separable state, $\rho=\rho_{A}\otimes \rho_{B}$, it is
obvious that,
\begin{eqnarray}
E_{ij}=\text{tr}(A_{i}\rho_{A})\text{tr}(B_{j}\rho_{B}).
\end{eqnarray}
Thus Eq.(\ref{equality}) surely should be satisfied for any
separable pure state.



We next consider a simple example of a two qubit entangled state,
$|\Psi\rangle=\sin\theta|00\rangle+\cos\theta e^{i\phi}|11\rangle$.
If Alice and Bob choose four measurement operators given above, then
we have,
\begin{eqnarray}
E_{11}E_{22}&=&\langle\sigma_{x}\sigma_{x}\rangle\langle\sigma_{y}\sigma_{y}\rangle=-\sin^{2}2\theta\cos^{2}\phi,\\
E_{12}E_{21}&=&\langle\sigma_{x}\sigma_{y}\rangle\langle\sigma_{y}\sigma_{x}\rangle=\sin^{2}2\theta\sin^{2}\phi
.
\end{eqnarray}
Equality holds only for, $\sin^{2}2\theta=0$, this means that any
pure entangled state violates (2). This is consistent with Gisin's
theorem \cite{gisin,gisinchen}. Here relation (2) can determine completely whether this
pure state is entangled or not.

Eq. (\ref{equality}) holds for any completely separable mixed state
$\rho =\rho _A\otimes \rho _B$, however, the general separable mixed
state does not satisfy this condition. Suppose two-qubit mixed
separable state have the form
$\rho=\sum_{m}p_{m}\rho_{A}^{m}\otimes \rho_{B}^{m}$, where
$\sum_{m}p_{m}=1$. This gives,
\begin{eqnarray}
E_{ij}E_{kl}=\sum_{mn}p_{m}p_{n}\text{tr}(A_{i}\rho_{A}^{m})
\text{tr}(B_{j}\rho_{B}^{m})\text{tr}(A_{k}\rho_{A}^{n})\text{tr}(B_{l}\rho_{B}^{n}),
\nonumber
\\
E_{il}E_{kj}=\sum_{mn}p_{m}p_{n}\text{tr}(A_{i}\rho_{A}^{m})\text{tr}(B_{l}
\rho_{B}^{m})\text{tr}(A_{k}\rho_{A}^{n})\text{tr}(B_{j}\rho_{B}^{n}).
\nonumber
\end{eqnarray}
Generally, equality (2) is violated by mixed separable state, we may
still need turn to Bell inequalities.


\section{Application to CV}
We next consider the application of
Eq. (\ref{equality}) to a CV system. Let us introduce the notations,
defining quadrature operators
$A_{1}=\hat{a}_{1}+\hat{a}_{1}^{\dag}$,
$A_{2}=i(\hat{a}_{1}-\hat{a}_{1}^{\dag})$ for Alice and
$B_{1}=\hat{a}_{2}+\hat{a}_{2}^{\dag}$,
$B_{2}=-i(\hat{a}_{2}-\hat{a}_{2}^{\dag})$ for Bob, where
$\hat{a}_{i}$ and $\hat{a}_{i}^{\dag}$ are boson annihilation and
creation operators. A simple example is that when system is prepared
in a superposition of single photon state,
$|\Psi\rangle=\sin\theta|01\rangle+\cos\theta e^{i\phi}|10\rangle$,
where we have used the standard Fock-state representation,
$|n\rangle _{A(B)}\equiv \frac {(\hat{a}_{1(2)}^{\dag})^n}{\sqrt
{n!}}|0\rangle _{A(B)}$, we obtain
$E_{11}E_{22}=-\sin^{2}2\theta\cos^{2}\phi$ and
$E_{12}E_{21}=\sin^{2}2\theta\sin^{2}\phi$. Eq.(\ref{equality}) is
only satisfied when the quantum state is separable.

We consider the CV system as the two modes squeeze process which
represents a nonlinear interaction of two quantized modes in a
nonlinear medium with a strong classical pump field. The entangled state
can be realized as the limiting case of a properly normalized two
mode squeezed state (TMSS) for infinitely large squeezing.
Explicitly, the TMSS
takes the form,
\begin{eqnarray}
|\text{TMSS}\rangle=e^{r(a_{1}^{\dag}a_{2}^{\dag}-a_{1}a_{2})}|00\rangle=\sum_{n=0}^{\infty}\frac{\tanh^{n}r}{\cosh
r}|nn\rangle . \label{tmss}
\end{eqnarray}
By using the quadrature operators defined above, we can calculate
that,
\begin{eqnarray}
E_{11}E_{22}&=&4\left(\sum_{n=0}^{\infty}\frac{\tanh^{2n+1}r}{\cosh^{2}r}(n+1)\right)^{2}
/\sum_{n=0}^{\infty}\frac{\tanh^{2n}r}{\cosh^{2}r},\nonumber \\
E_{12}E_{21}&=&0.
\end{eqnarray}
Eq.(\ref{equality}) is not satisfied regardless of squeezing
parameter $r, (r>0)$. On the other hand,
we already know, the TMSS violates the
Bell inequality for any $r, (r>0)$, see for example Ref.\cite{Chen}. Of course, the
degree of violation is still related with $r$. Thus
Eq.(\ref{equality}) can be used to detect whether the involved state
is entangled or not for both discrete and CV systems.



Though the proposed relation is not satisfied by general mixed state, it is,
however, can be applied for case of mixed state constructed by a
pure state with depolarizing noise. Let
us start from the single photon entangled state,
$|\Psi\rangle=\sin\theta|01\rangle+\cos\theta e^{i\phi}|10\rangle$.
Taking the interference of white-noise like decoherence into
account, the system is now in state,
\begin{eqnarray}
\rho=p|\Psi\rangle\langle\Psi|+(1-p)\frac{I}{\text{tr}I},
\end{eqnarray} 
where $p$ is the probability that the state is unaffected by noise. We
find $E_{11}E_{22}=-p^{2}\sin^{2}2\theta\cos^{2}\phi$ and
$E_{12}E_{21}=p^{2}\sin^{2}2\theta\sin^{2}\phi$. Then $p^{2}$
becomes an overall factor in both side of equality (2). If the
degree of violation is defined by the ratio of r.h.s. and l.h.s. of
Eq.(\ref{equality}), it is unaffected by this depolarizing noise. Note that when
$p$ becomes small, the state itself becomes separable \cite{Werner}.
Similar as for case of TMSS in (\ref{tmss}), we consider state,
$\rho =p|TMSS\rangle \langle TMSS|+(1-p)\frac{I}{\text{tr}I}$.
By straightforward calculation,  we can also find $p^{2}$ as an overall
factor in both sides of Eq.(\ref{equality}).

The importance of this property can be as follows.
We know that the main loophole in experimental observation of
Bell inequality is caused by the
detection loss. Also it is known that the photon-detection with efficiency
$\eta$ is practically equivalent to the ideal detection after the
signal $\hat{a}$ is mixed with a vacuum $\hat{v}$ at a beam splitter
of transmissivity $\sqrt{\eta}$. Namely, the observed signal
$\hat{a}_{o}$ is expressed by
$\hat{a}_{o}=\sqrt{\eta}\hat{a}+\sqrt{1-\eta}\hat{v}$. This gives
$E_{11}E_{22}=-\eta^{2}\sin^{2}2\theta\cos^{2}\phi$ and
$E_{12}E_{21}=\eta^{2}\sin^{2}2\theta\sin^{2}\phi$ for single photon
entangled state $|\Psi\rangle=\sin\theta|01\rangle+\cos\theta
e^{i\phi}|10\rangle$. Thus by using the balanced homodyne detection to
measure the quadrature amplitudes at each mode, $\eta^{2}$
becomes an overall factor in both sides, which makes our scheme
insensitive to detector efficiency $\eta $. Thus the relation (2)
can detect the entangled state of CV effectively.

\section{Generalization for multipartite state and CFRD inequality}
The correlation function equality (2) can be generalized to multipartite
system straightforwardly by increasing the number of parties.
Consider $n$ spatially separated observers and each of them can choose between
$m$ experimental settings. The $i_{j}$-th measurement of party $j$ is
denoted by $M_{j}^{i_{j}}$, $j=1,...,n, i_{j}=1,...,m$. The
correlation function equality with respect to this system can be
taken as,
\begin{align}
E_{i_{1}...i_{n}}E_{i_{1}^{\prime}...i_{n}^{\prime}}=P
\{E_{i_{1}...i_{n}}E_{i_{1}^{\prime}...i_{n}^{\prime}}\},
\label{multi}
\end{align}
where $E_{i_{1}...i_{n}}=\langle
M_{1}^{i_{1}}...M_{n}^{i_{n}}\rangle$ and $P$ represent the exchange
of any number of measurement of observables between two correlation
function $i_{1}...i_{n}:i_{1}^{\prime}...i_{n}^{\prime}$. Equality
(\ref{multi}) can easily be demonstrated to be true for separable pure state.
We remark that, in general, we don't have to restrict the number of
correlation function to be two in both sides.
The simplest one of all those equalities (\ref{multi}) is $\langle
A_{i}B_{j}\rangle=\langle A_{i}\rangle\langle B_{j}\rangle$, which
can be obtained by choosing $A_{k}=I$ and $B_{l}=I$ in (2). This equality
can also be expressed as $p(A_{i}=a,B_{j}=b)=p(A_{i}=a)p(B_{j}=b)$,
where $a$ and $b$ are outcomes of Alice's and Bob's measurement. The
failure of this equality in entangled pure state suggest that
local measurement may act as a detection for the entangled states.

It is obvious that pure separable state for multipartite
system does not violate this equality. While quantum entangled state
will have a violation. As an example, we assume each one
of $n$ parties can choose between two measurements, $i_{j}=1,2$. $P$
represents exchange first $r$ measurement between two correlation
function, which is
\begin{align}
P\{E_{i_{1}\ldots i_{n}}E_{i_{1}^{\prime}\ldots i_{n}^{\prime}}\}=
E_{i_{1}^{\prime}\ldots i_{r}^{\prime}i_{r+1}\ldots i_{n}}
E_{i_{1}\ldots i_{r}i_{r+1}^{\prime}\ldots i_{n}^{\prime}}.
\label{commut}
\end{align}
The violation can be derived immediately when system is prepared in
entangled state. We define quadrature operators \cite{CFRD},
\begin{equation}
\begin{split}
M_{j}^{1}&=\hat{a}_{j}+\hat{a}_{j}^{\dag},\\
M_{j}^{2}&=\hat{a}_{j}e^{-i\phi_{j}}+\hat{a}_{j}^{\dag}e^{i\phi_{j}},
\end{split}
\end{equation}
where $\phi_{j}$ is local measurement phase parameters with respect
to observer $j$. If the first $r$ observers set their local phase
parameters are $+\pi/2$, and the remaining set theirs as $-\pi/2$.
Measurement is performed in quantum state,
\begin{eqnarray}
|\Psi\rangle=\sin\theta|0\rangle^{r}|1\rangle^{n-r}+\cos\theta
e^{i\phi}|1\rangle^{r}|0\rangle^{n-r} ,
\label{state}
\end{eqnarray}
which is the superposition of
$r$ modes ($n-r$ modes) all occupying one (no) photon and vice
versa. Then we find,
\begin{eqnarray}
E_{1\cdots1}E_{2\cdots2}&=&\sin^{2}2\theta\cos\phi\cos(\phi-n\pi/2),\\
E_{1\cdots12\cdots2}E_{2\cdots21\cdots1}&=&\sin^{2}2\theta \cos[\phi-(n-r)
\pi/2]
\nonumber \\
&&\times \cos(\phi-r\pi/2).
\end{eqnarray}
Choose these parameters appropriately, one can always find
that two sides of Eq. (\ref{multi}) does not equal. When $n=2$
and $r=1$,  we  recover the previous result. We remark that like the $n=2$ case,
general mixed separable state does not always satisfy the equality.

We now study the CFRD inequality, which is a Bell-type inequality for
continuous variables with respect to $n$ parties and each party can
carry out two dichotomic measurements \cite{CFRD}. It takes the form,
\begin{eqnarray}
\left|\left\langle\prod_{j=1}^{n}\left[M_{j}^{1}+iM_{j}^{2}\right]\right\rangle\right|^{2}\leq
\left\langle\prod_{j=1}^{n}\left[(M_{j}^{1})^{2}+(M_{j}^{2})^{2}\right]\right\rangle,
\label{cfrd}
\end{eqnarray}
where $M_{j}^{i_{j}}$ are $i_{j}$-th measurement choice with respect
to observer $j$ as denoted before.

First let us consider the simplest $n=2$ case.
To derive the inequality, it is assumed that,
$E_{11}E_{22}=E_{12}E_{21}$,
which is exact the
condition (2),
where $E_{i_{1}i_{2}}^{2}=\left\langle
M_{1}^{i_{1}}M_{2}^{i_{2}}\right\rangle^{2}$,see \cite{CFRD}.
So the CFRD inequality is based on the fact that all
measurement have nonnegative variance, that is,
\begin{eqnarray}
\left\langle
(M_{1}^{i_{1}})^{2}(M_{2}^{i_{2}})^{2}\right\rangle -\left\langle M_{1}^{i_{1}}M_{2}^{i_{2}}\right\rangle^{2}\ge 0.
\label{inequal}
\end{eqnarray}
Thus the CFRD inequality (\ref{cfrd}) can be violated only
for case that (2) is not satisfied. Thus, the violation of
equation, $E_{11}E_{22}\not =E_{12}E_{21}$, is more powerful than
the violation of CFRD inequality to detect the entangled state. As we have shown,
this property can detect all entangled state of form (\ref{state}).

The CFRD inequality is proposed in Ref.\cite{CFRD}, they found that
the inequality is violated for case $n\ge 10$ for state (\ref{state}).
Then in Ref.\cite{Grassl1}, the violation condition is extended to case
$n\ge 5$. Additionally if we apply the method of \cite{Chen}, the violation
case can be extended to $n\ge 3$, see next section for detail.
On the other hand, in this paper, we find the violation of Eq.(2) can detect
all entangled state of form (\ref{state}).
Thus we show that the violation of Eq.(2) is
a necessary condition for the violation of CFRD inequality.
We remark that the entangled state can be detected  by a different approach
by using two classes of multimode Bell inequalities
proposed in Ref.\cite{Grassl2} for case $n\ge 2$. Here we present
a much simpler method which can detect all cases $n\ge 2$.



The general case seems straightforward. For simplicity, we denote
$M=\left\langle\prod_{j=1}^{n}[M_{j}^{1}+iM_{j}^{2}]\right\rangle$.
The real part of $M$ is the sum of $n$ partite correlation function
which only contain even numbers of measurement observable with
$i_{j}=2$ and the imaginary part of $M$ contain odd numbers. We always
find that cross term of correlation function appears in
$(\text{Re}M)^{2}$ and its opposite counterpart which generally have
form (\ref{commut}) appears in $(\text{Im}M)^{2}$. For separable state pure
state,
they vanish which leave us with terms of all square of correlation
function of $n$ parties, which is
$M=M^{\prime}\equiv\sum_{i_{1},\cdots,i_{n}=1,2}\langle
M_{1}^{i_{1}}\cdots M_{n}^{i_{n}}\rangle^{2}$. This is always less
than or equal to $\sum_{i_{1},\cdots,i_{n}=1,2}\langle
(M_{1}^{i_{1}})^{2}\cdots (M_{n}^{i_{n}})^{2}\rangle$.
As has been demonstrated in $n=2$ case,  the fact $M^{\prime}$ is
always less than or equal to the right side of (\ref{cfrd}) leads us
to know, as expected, all the violation of CFRD inequality comes
from the violation of Eq.(2) and its generalization.




\section{Violation of CFRD inequality}
Here we show that the violation of CFRD inequality
can be extended to case $n\ge 3$ for state (\ref{state}).
We introduce the notations,
\begin{eqnarray}
s_z&=&\sum _{k=0}^\infty [|2k+1\rangle \langle 2k+1|-|2k\rangle \langle 2k|],
\\
s_-&=&\sum _{k=0}^{\infty }|2k\rangle \langle 2k+1|=(s_+)^{\dagger }.
\end{eqnarray}
Those operators can be used
to find the violation of CHSH inequality for TMSS (\ref{tmss}), see Ref.\cite{Chen}.
The measurement operators in (\ref{cfrd}) can be defined as,
\begin{eqnarray}
M_j^1&=&s_-^j+s_+^j,\\
M_j^2&=&s_-^je^{-i\phi _j}+s_+^je^{i\phi _j}.
\end{eqnarray}
For state (\ref{state}),
similar as in Ref.\cite{CFRD},
let the first $r$ phase parameters $\phi _j=\pi /2$, the last $n-r$ phase parameters $\phi _j=-\pi /2$. So for
$j\le r$, $M_j^1+iM_j^2=2s_-^j$, and for $j>r$, $M_j^1+iM_j^2=2s_+^j$, also we have $(M_j^1)^2+(M_j^2)^2=
2(s_-^js_+^j+s_+^js_-^j)$. Now let us consider CFRD inequality (\ref{cfrd}), we find,
\begin{eqnarray}
l.h.s&=&2^{2n}|\langle s_-^1...s_-^rs_+^{r+1}...s_+^n\rangle |^2,
\\
r.h.s&=&2^n\langle (s_-^1s_+^1+s_+^1s_-^1)...(s_-^ns_+^n+s_+^ns_-^n) \rangle .
\end{eqnarray}
With the exact form of (\ref{state}), we find,
\begin{eqnarray}
l.h.s&=&2^{2n}\cos ^2\theta \sin ^2\theta ,
\\
r.h.s&=&2^n.
\end{eqnarray}
We then let
$\cos ^2\theta =\sin ^2\theta =1/2$, the $l.h.s$ achieves the
maximal point $2^{2n}/4$. Then one may find that when $n=3$,
\begin{eqnarray}
l.h.s=16>r.h.s=8,
\end{eqnarray}
The CFRD inequality (\ref{cfrd}) is violated, 
and the violation is true for cases $n>3$. Then here
the violation region is for $n\ge 3$ which extends the results
in Refs.\cite{CFRD,Grassl1} in which the violation cases are:
$n\ge 10$ and $n\ge 5$, respectively.

\section{Conclusion}
Much of recent study of entanglement focuses on
Bell type inequalities, both theoretically and experimentally.
In this paper, we propose a simple equality by local measurement 
to detect the entangled state.
It provides a much simpler criterion to detect entangled
state for discrete and CV systems, additionally it may have the
advantage that the violation is not affected by depolarizing noise
of decoherence. In addition, it is the origination of the violation of CFRD
inequality. Actually, all entangled state can be detected by this condition.
Besides of CV systems of optics, it can also
be easily checked in some other physical
systems, for example, in superconducting systems,
see Ref. \cite{youjq} for a review.
The proposed equation is simple and powerful, and is an important complement to
Bell inequalities. Depending on question and system, we may choose to
find the violation of either Bell inequalities or this simple equation.

This work is supported by NSFC(10931006, 10975180, 10974247,
11075126),``973'' program (2010CB922904).

\end{document}